\documentclass[12pt]{article}
\usepackage[latin1]{inputenc}
\usepackage{amsmath}
\usepackage{amsthm}
\usepackage{amssymb}
\usepackage{psfrag}
\usepackage{graphicx}
\usepackage{url}

\newtheorem{thm}{Theorem}
\newtheorem{lem}[thm]{Lemma}

\newcommand{\nlb}[1]{mod{}$#1${}NLB}

\begin{document}

\title{No nonlocal box is universal}
\author{Fr\'ed\'eric Dupuis$^{\star}$ \quad Nicolas Gisin$^{\dagger}$ \quad Avinatan Hasidim$^{\S}$ \\ Andr\'e Allan M\'ethot$^{\dagger}$ \quad Haran Pilpel$^{\ddag}$\\[0.5cm]
\normalsize\sl $^{\star}$ D\'epartement d'informatique et de recherche op\'erationnelle\\ \normalsize\sl Universit\'e de Montr\'eal, Montr\'eal, Qu\'ebec, \textsc{Canada}\\[-0.1cm]
\normalsize\url{dupuisf}\textsf{@}\url{iro.umontreal.ca}\\[0.5cm]
\normalsize\sl $^{\dagger}$ Group of Applied Physics, Universit\'e de Gen\`eve, Gen\`eve, \textsc{Switzerland}\\[-0.1cm]
\normalsize\url{{nicolas.gisin,andre.methot}}\textsf{@}\url{physics.unige.ch}\\[0.5cm]
\normalsize\sl $^{\S}$ Department of Computer Science, Hebrew University, Jerusalem, \textsc{Israel}\\[-0.1cm]
\normalsize\url{avinatanh@gmail.com}\\[0.5cm]
\normalsize\sl $^{\ddag}$ Einstein Institute of Mathematics, Hebrew University, Jerusalem, \textsc{Israel}\\[-0.1cm]
\normalsize\url{haranp@math.huji.ac.il} }

\maketitle

\begin{abstract}
We show that standard nonlocal boxes, also known as Popescu-Rohrlich
machines, are not sufficient to simulate any nonlocal correlations
that do not allow signalling. This was known in the multipartite
scenario, but we extend the result to the bipartite case. We then
generalize this result further by showing that \emph{no} finite set
containing any finite-output-alphabet nonlocal boxes can be a
universal set for nonlocality.
\end{abstract}

\section{Introduction}\label{sec:intro}
\begin{figure}[thb!]
\centering
\psfrag{x}{$x \in \{0,1\}$}
\psfrag{y}{$y \in \{0,1\}$}
\psfrag{a}{$a \in_r \{0,1\}$}
\psfrag{b}{$b \in_r \{0,1\}$}
\psfrag{Alice}{Alice}
\psfrag{Bob}{Bob}
\psfrag{eqn}{$b-a \mod 2 = xy$}
\includegraphics{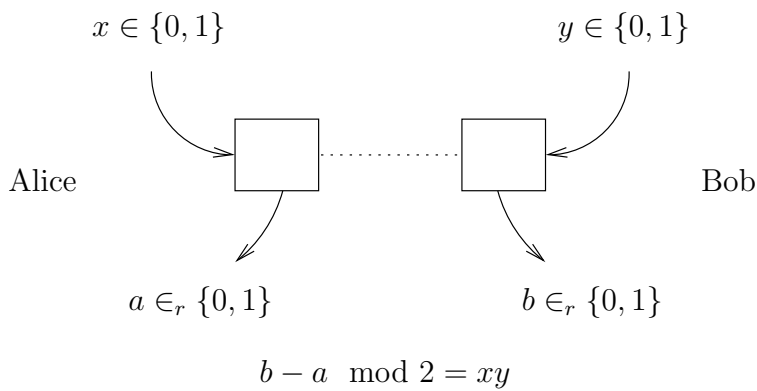}
\caption{The standard nonlocal box. The notation $a \in_r \{0,1\}$
means that $a$ is uniformly distributed over $\{0,1\}$.}
\label{standard-nlb}
\end{figure}

Nonlocality refers to a multi-party process that, while it does not
allow for communication, would classically necessitate communication
for the different parties to perform. One classic example is the
following ``nonlocal box'' (see Figure~\ref{standard-nlb}): imagine
that two parties, henceforth referred to as Alice and Bob, have a
black box into which they can each enter one bit of their choice and
the box gives each of them a random bit such that the exclusive-or
of the output bits is equal to the AND of the input bits~\cite{PR2}.
If one attempts to implement this box without communication in a
classical world, it is easy to show that it is impossible to succeed
more than 75\% of the time~\cite{CHSH}. On the other hand, if the
output bits are always uniformly distributed for all inputs, then it
is clear that this box does not permit communication between Alice
and Bob, since the local probability distribution does not depend on
the parties inputs.

\begin{figure}[thb!]
\centering
\psfrag{x}{$x\in \mathcal{X}$}
\psfrag{y}{$y\in \mathcal{Y}$}
\psfrag{a}{$a\in \mathcal{A}$}
\psfrag{b}{$b\in \mathcal{B}$}
\psfrag{Alice}{Alice}
\psfrag{Bob}{Bob}
\includegraphics{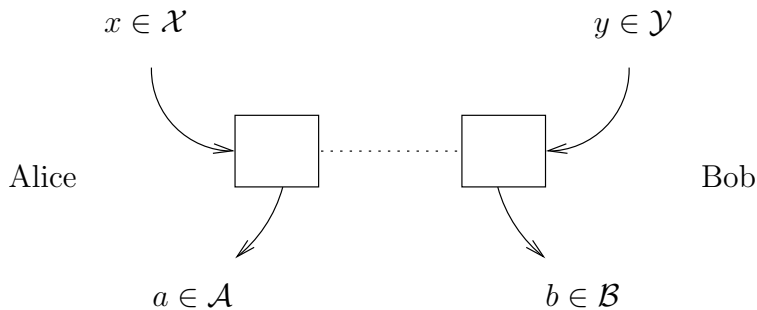}
\caption{A generic nonlocal box. Alice and Bob input $x$ and $y$
respectively, and receive $a$ and $b$ respectively. The resulting
probability distribution $p(a,b|x,y)$ cannot be reproduced
classically without communication, and yet does not itself allow
Alice and Bob to communicate.} \label{generic-nlb}
\end{figure}

We shall call this box the mod2NLB for reasons that will become
obvious later. A general nonlocal box is a device such that: given
inputs $x$ from Alice and $y$ from Bob, they output values $a$ and
$b$ such that the resulting probability distribution $p(a,b|x,y)$
cannot be reproduced classically without communication, yet cannot
itself be used to communicate (see Figure~\ref{generic-nlb}).

The original motivation for studying nonlocality is quantum
mechanics. Indeed, in quantum mechanics, entanglement allows us to
achieve nonlocal correlations similar to those described above. John
Bell \cite{bell} was the first to show that measurement on shared
entangled quantum state could produce correlations that cannot be
locally simulated classically. Later, Clauser, Horne, Shimony and
Holt \cite{CHSH} came up with an inequality (the so-called CHSH
inequality) that provided a condition for a certain type of
correlations to be explainable by classical means alone, and showed
that quantum mechanics violated it in some cases. The mod2NLB was
directly inspired by this inequality: the mod2NLB violates the CHSH
inequality to its maximal algebraic value while quantum mechanics
can be used to simulate a mod2NLB with up to approximately 85\%
efficiency~\cite{cirelson80}.

The nonlocality of quantum mechanics has been known for a long time,
but has only recently started to be studied by itself, i.e.\
independently from the study of entanglement. It is hoped that such
an independent study will allow us to understand the implications of
nonlocality in quantum mechanics more thoroughly. Furthermore, there
is proof that entanglement and nonlocality are not the same. The
first example came from~\cite{bgs05}, where it was proved that a
single mod2NLB is not sufficient to simulate a non-maximally
entangled pair of qubits, even though a perfect simulation of all
correlations of the maximally entangled state of two qubits is
possible with only one mod2NLB\cite{cgmp05}. The final proof that
entanglement and nonlocality are different resources came
in~\cite{bm06}, where it was proven that a simulation of $n$
maximally entangled pair of qubits required $\Omega(2^n)$ mod2NLBs.

This asserts that entanglement and nonlocality should be treated as
different types of resources. But while we know a fair bit about
entanglement, comparatively speaking little is known about
nonlocality. For instance, we have been able to isolate the
maximally entangled state of two qubits as the ``unit'' of bipartite
entanglement, since, together with local operations and classical
communication, it allows us to create any other entangled state,
provided we have enough copies. Is there an analogous concept for
nonlocality? Would it be possible to identify a similar ``unit of
nonlocality'', that would allow us to create other bipartite
nonlocal correlations? The mod2NLB was the obvious candidate: its
minimal size (binary inputs and outputs) and the fact that it
violates the CHSH inequality, the only nontrivial inequality at
these dimensions, maximally made it very attractive from that point
of view.

There are more encouraging signs to support the mod2NLB's claim as
the universal resource of nonlocality. One particularly interesting
result is that the mod2NLB makes communication
complexity~\cite{kn97} trivial~\cite{vandamCC,cleveCC}. That is, if
two players are allowed to use mod2NLBs, they can compute any
boolean function of their inputs with a single bit of communication,
regardless of what the function is.

In light of these facts, it is tempting to think that the mod2NLB
could be considered as a unit of nonlocality that can be used to
generate any other bipartite nonlocal correlation. Some progress has
been made in this direction: in~\cite{BP}, Barrett and Pironio have
shown that mod2NLBs alone can be used to simulate any two-output
bipartite boxes. However, they have shown that there exist
multipartite nonlocal correlations that cannot be simulated by
mod2NLBs alone. What about the bipartite scenario? In~\cite{BLMPPR},
a family of bipartite nonlocal boxes is presented which can generate
every two-input bipartite box. In this paper, we present a
complementary negative result: we show that no \emph{finite} set
containing any general bipartite nonlocal boxes can simulate all
bipartite nonlocal boxes.

We start, for intuition, by proving the non-universality of the
traditional nonlocal box in Section~\ref{sec:mod2nlb}. We prove that
a finite number of mod2NLB cannot perfectly simulate the mod3NLB, to
be defined at the beginning of Section~\ref{sec:mod2nlb}. In
Section~\ref{sec:modpnlb}, we generalize the result by proving that
no finite set of finite-output-alphabet nonlocal boxes can be
universal. We then conclude in Section~\ref{sec:conclusion}.

\section{The non-universality of the traditional nonlocal box}\label{sec:mod2nlb}

\begin{figure}[thb!]
\centering
\psfrag{x}{$x \in \{0,1\}$}
\psfrag{y}{$y \in \{0,1\}$}
\psfrag{a}{$a \in_r \{0,1,\dots,p-1\}$}
\psfrag{b}{$b \in_r \{0,1,\dots,p-1\}$}
\psfrag{Alice}{Alice}
\psfrag{Bob}{Bob}
\psfrag{eqn}{$b-a \mod p = xy$}
\includegraphics{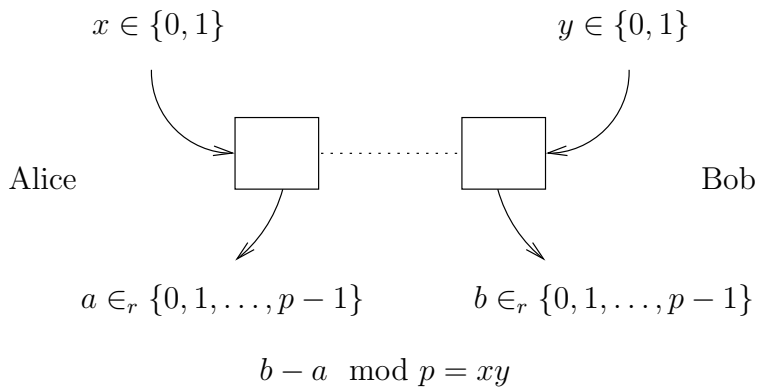}
\caption{A graphical description of the mod$p$NLB.}\label{modpfig}
\end{figure}

We will first begin by introducing the mod3NLB: $x \in \{0,1\}$, $y
\in \{0,1\}$, $a \in \{0,1,2\}$, $b \in \{0,1,2\}$, and
\begin{equation}
    \label{boxA3}
    p(a,b|x,y) = \left\{
    \begin{array}{ll}
        \frac{1}{3} & \hbox{ if } b-a = xy \mod 3\\
        0 & \hbox{ otherwise}
    \end{array}
    \right.
\end{equation}
See Figure~\ref{modpfig} for a graphical representation of the
general mod$p$NLB. Clearly, this does not allow communication
between Alice and Bob, since, taken alone, $a$ is completely
independent from $x$ and $y$, and likewise for $b$. The mod3NLB is
therefore a valid nonlocal box and a simple extension of the
traditional mod2NLB. It would seem reasonable, especially in light
of~\cite{BP}, that such a nonlocal box could be simulated by
mod2NLBs. However, the following theorem states the opposite.

\begin{thm}\label{thm:mod2NLB}
It is impossible to simulate the mod3NLB exactly using a finite
number of mod2NLBs, infinite shared randomness and no communication
between the two players.
\end{thm}
\begin{proof}
Let's assume that there exists an algorithm (which may be
probabilistic) that can perfectly simulate one instance of the
mod3NLB using $N$ mod2NLBs; we will then show that this assumption
leads to a contradiction.

First, we can reduce the problem to deterministic algorithms in the
following manner: any probabilistic algorithm can be represented as
a collection of deterministic algorithms $\alpha_i$, each with a
certain probability of being selected. Since we require perfect
simulation of the mod3NLB, the outputs of the algorithm must satisfy
the equation $b-a = xy$ with probability 1; hence each algorithm
$\alpha_i$ with nonzero probability in any probabilistic algorithm
must also satisfy this equation with probability 1. For our
contradiction, we can therefore restrict ourselves to deterministic
algorithms, since a correct probabilistic algorithm exists only if a
deterministic algorithm satisfying $b-a = xy$ exists.

Observe first that, for all deterministic algorithms, the output $a$
is completely determined by $x$ and the $N$ output bits that Alice
got from the mod2NLBs, since we can simulate Alice's algorithm using
only those $N+1$ values. Likewise, we can do this on Bob's side to
determine $b$ from $y$ and Bob's mod2NLB outputs. To formalize this,
let $z_A$ be the bit-string that Alice obtained from the mod2NLBs,
and $z_B$ be Bob's bit-string. Then there exist two functions $F_A$
and $F_B$ such that $a = F_A(x, z_A)$ and $b = F_B(y, z_B)$. Note
that $z_A$ and $z_B$ are uniformly distributed on $\{0,1\}^N$. We
can now define the following two probability distributions:
\begin{eqnarray}
p_A(a|x) &=& \Pr\{F_A(x,Z) = a\}\\
p_B(b|y) &=& \Pr\{F_B(y,Z) = b\}
\end{eqnarray}
where $Z$ is a random variable uniformly distributed on $\{0,1\}^N$.

Let us note that $2^N$ is not divisible by 3, therefore $p_A$ and
$p_B$ cannot be uniform for any value of $x$ and $y$. Since we must
be able to simulate the box perfectly, we must at least have:
\begin{eqnarray}
    \label{cdn1} p_A(q|0) &=& p_B(q|0)\\
    \label{cdn2} p_A(q|0) &=& p_B(q|1)\\
    \label{cdn3} p_A(q|1) &=& p_B(q|0)\\
    \label{cdn4} p_A(q|1) &=& p_B(q+1|1)
\end{eqnarray}
where additions are performed mod 3. Condition $(\ref{cdn1})$ comes
from the fact that if $x=y=0$, then $b=a$ every time, hence the two
marginal distributions must be identical, and therefore $p_A(q|0) =
p_B(q|0)$. The other three conditions correspond to similar
conditions when the inputs are $(a,b) = (0,1)$, $(a,b) = (1,0)$ and
$(a,b) = (1,1)$ respectively.

These conditions lead to a contradiction: $(\ref{cdn1})$ and
$(\ref{cdn3})$ imply that $p_A(q|0) = p_A(q|1)$, which means that
$(\ref{cdn2})$ and $(\ref{cdn4})$ imply that $p_B(q|1) =
p_B(q+1|1)$. Since $p_B(q|1)$ cannot be uniform, we are forced to
conclude that perfect simulation of the mod3NLB with $N$ mod2NLBs is
impossible.
\end{proof}

\section{Generalization to a finite set of nonlocal boxes}\label{sec:modpnlb}

The result of Section~\ref{sec:mod2nlb} can be generalized to a finite set of nonlocal boxes, as defined in Section~\ref{sec:intro} and represented in Figure~\ref{generic-nlb}, where the dimensions of the output sets are finite, i.e.\ $\vert \mathcal{A}\vert , \vert \mathcal{B} \vert < \infty$. Before turning to the main theorem and its proof, we need to define the mod$p$NLB in the following manner:
\begin{equation}\label{boxAn}
p(a,b|x,y) = \begin{cases}
        \frac{1}{p} \text{\quad if } b-a = xy \mod p\\
        0 \text{\quad otherwise}
    \end{cases},
\end{equation}
where $\vert \mathcal{X}\vert = \vert \mathcal{Y} \vert =2$ and $\vert \mathcal{A}\vert = \vert \mathcal{B} \vert =p$. This family of nonlocal boxes was first defined in~\cite{BLMPPR}. It was also shown that this family, which is an infinite set, could be used to simulate any two-input bipartite nonlocal box. They also showed that a mod$p$NLB and a mod$q$NLB could simulate a mod$r$NLB, where $r=pq$. Here, we shall prove that for any finite set of nonlocal boxes, whatever their nature, there exist a member of this family that cannot be simulated by a finite number of boxes from the set. Before turning to our main theorem and its proof, we first need two technical lemmas.
\begin{lem}\label{lem:finite}
If a first-order formula $\phi$ is true in $\mathbb{C}$, it is then true in another field $\mathfrak{F}_q$ which has a large enough characteristic $q$.
\end{lem}
\begin{proof}
	Recall that the characteristic $k$ of a field $\mathfrak{F}$ is the smallest positive integer such that for every $x \in \mathfrak{F}$, $x+\cdots+x = 0$ where $x$ is summed $k$ times; if no such $k$ exists, the characteristic of $\mathfrak{F}$ is zero by definition. It is a known result in mathematical logic that the theory of algebraically closed fields with specified characteristic is complete (see, e.g., \cite{bell-slomson}, pp. 178--179). This means that given a set of axioms which define an algebraically closed field of characteristic $k$, then every first-order formula can either be proven true or false from these axioms alone. Thus, since $\mathbb{C}$ is an algebraically closed field of characteristic 0, if a first-order formula is true in $\mathbb{C}$, it must be provable in the theory of algebraically closed fields of characteristic 0.

Now, the axioms of the theory of an algebraically closed field of characteristic $0$ consist of two disjoint sets:
\begin{itemize}
\item A set of axioms $S_{\textrm{alb}}$ which define an algebraically closed field
\item A set of axioms $S_{\textrm{chr}} = \{\tau_p : p \mbox{ prime}\}$ where $\tau_p$ is the axiom that the field is not of characteristic $p$.
\end{itemize}
The proof of $\phi$ in this theory is finite, and thus uses a finite subset of axioms $S$. Specifically it only uses a finite subset of $S_{\textrm{chr}}$. Let $q$ be a prime large enough such that $\tau_{p} \notin S$ for every $p \geq q$, and let $\mathfrak{F}_q$ be any algebraically closed field of characteristic $q$. The proof only uses axioms which hold in $\mathfrak{F}_q$ and, therefore, the statement $\phi$ is true in $\mathfrak{F}_q$.
\end{proof}
\begin{lem}\label{lem:extension}
No finitely generated extension ring of $\mathbb{Z}$ contains all numbers of the form $1/p$, where $p$ is a prime number.
\end{lem}
\begin{proof}
Let $T$ be some extension ring of $\mathbb{Z}$ generated by a finite set $G = \{ q_1, \ldots, q_n \} \subset \mathbb{R}$ over $\mathbb{Z}$, meaning that $T$ is composed of numbers which are the sum of products of numbers in $G$ and $\mathbb{Z}$. Assume, by contradiction, that all numbers of the form $1/p$, where $p$ is prime, are contained in $T$. Let $\mathfrak{I}$ be the set of all $n$-variable polynomial relations with coefficients in $\mathbb{Z}$ satisfied by $q_1, \ldots, q_n$; it can be shown that $\mathfrak{I}$ is an \emph{ideal}. From the Hilbert basis theorem \cite{hilbert-basis}, it follows that $\mathfrak{I}$ is finitely generated; this means that there exists a set of polynomials $F = \{ f_1, \ldots, f_m \}$ such that every $P$ in $\mathfrak{I}$ can be written as $P = \sum_{k=1}^{m} r_k f_k$, where $\left\{ r_1,\ldots,r_m \right\}$ is a set of polynomials that depends on $P$.

Now, consider the following first-order formula;
\begin{equation}
\phi :\ \exists y_1, \ldots, y_n [ f_1(y_1, \ldots, y_n) = \cdots =
f_m(y_1, \ldots, y_n) = 0 ].
\end{equation}
Obviously, $\phi$ must be true in $\mathbb{R}$, since $q_1,\ldots,q_n$ satisfies the formula. It must therefore also be true in $\mathbb{C}$ and, by Lemma \ref{lem:finite}, in a field $\mathfrak{F}_p$ of sufficiently large characteristic $p$.

Let $y_1, \ldots, y_n$ be the elements of $\mathfrak{F}_p$ guaranteed by $\phi$, which satisfy $f_1, \ldots, f_m$. By hypothesis, $1/p \in T$, so $1/p$ can be expressed as a sum of products of numbers in $G$. Thus, there is some $n$-variable polynomial $g$ with integer coefficients such that $g(q_1,\ldots,q_n) = 1/p$ and therefore $1 - pg(q_1,\ldots,q_n)=0$. This means that $1 - pg \in \mathfrak{I}$. We then have that $1-pg(y_1,\ldots,y_n) = 0$ in $\mathfrak{F}_p$. But in $\mathfrak{F}_p$, $p=0$ by definition and $1-pg(y_1,\ldots,y_n) = 1$. Thus we have a contradiction.
\end{proof}
\begin{thm}\label{thm:modpNLB}
Let $S$ be a finite set of nonlocal boxes, each with a finite output
alphabets $\mathcal{A}$ and $\mathcal{B}$. Then there exists $p$
such that the \nlb{p} cannot be simulated by a finite number of
nonlocal boxes taken from the set $S$.
\end{thm}
\begin{proof}
Let us first note that, as in the Proof of
Theorem~\ref{thm:mod2NLB}, the fact that the relation of the \nlb{p}
must be perfectly simulated entails that we can consider only
deterministic protocols; meaning that the only source of randomness
is the nonlocal boxes from the set $S$, that we do not have any
shared random variables and that the functions $F_A$ and $F_B$, with
which we calculate the final output of our simulation from the
internal nonlocal boxes and the input, are deterministic. Let
$p_k^{(l)}$ be the marginal probability of Alice's output $k$ of the
nonlocal box $l$ from the set $S$. Likewise, let $q_k^{(l)}$ be the
marginal probability of Bob's output $k$ of the nonlocal box $l$
from the set $S$. Let $P = \{ p_1^{(1)}, p_2^{(1)}, \ldots ,
p_1^{(2)}, \ldots \}$ and $Q = \{ q_1^{(1)}, q_2^{(1)}, \ldots ,
q_1^{(2)}, \ldots \}$ be the total collection of probabilities for
outputs of $S$. $P$ and $Q$ are obviously finite, since each output
alphabet and $S$ are finite.

Let $a$ be some element in the \nlb{p} output alphabet. From the
fact that the output of the simulation is calculated from a
deterministic function on the input and the output of the internal
nonlocal boxes, the marginal probability of $a$ must be some linear
combination (with integral coefficients) of products of
probabilities in the original set. Mathematically, $a$ must be an
element of the integral extension ring generated by $P$ over
$\mathbb{Z}$. In other words, $a$ is in the ``set'' of numbers
generated the addition of products of numbers from $P$ and
$\mathbb{Z}$.

However, it is proven in Lemma~\ref{lem:extension} that $\mathbb{Q}$
is not a finitely generated extension ring of $\mathbb{Z}$. Meaning
that it is impossible to create every number in $\mathbb{Q}$ by the
addition of products of numbers from $P$ and $\mathbb{Z}$.
Therefore, there exists some $q = 1/p \in \mathbb{Q}$ which is not
in the integral extension ring generated by $P$ over $\mathbb{Z}$.
Hence, the \nlb{p} cannot be simulated by a finite number of
nonlocal boxes from the set $S$.
\end{proof}

\section{Discussion and Conclusion}\label{sec:conclusion}

We have proven that no finite set of finite-output-alphabet nonlocal
boxes can be universal. Therefore only nonlocal boxes with an
infinite number of outputs can be. However, there exists no
compelling candidates, and one might argue that such a box would be
even more artificial and less elegant than the traditional mod2NLB.

It is to be noted that our result does not contradict those
of~\cite{BP}, since the universality of the family of mod$p$NLBs is
defined for binary input nonlocal boxes and requires an infinite set
of mod$p$NLBs.

Our result exhibits yet a new difference between nonlocality and
entanglement. The latter has a very simple and attractive universal
resource, the maximally entangled pair of qubits, while the former
has no such things. Our result also suggest that one must be careful
about statements made with the traditional mod2NLB, for it cannot be
associated with a general idea of nonlocality. It is important to
stress at this point that we do not believe research in nonlocal
boxes to be futile. For example, one can still uncover some
intuitions about Nature when studying the mod2NLB.
In~\cite{bblmtu05}, it was proven, using mod2NLBs, that if quantum
mechanics were slightly more nonlocal, it would have drastic and
arguably unbelievable consequences in communication complexity.

This work raises a philosophical question. What is the difference
between entanglement and nonlocality? Why does entanglement have a
universal resource while nonlocality doesn't? It is tempting to
think that it might be related to the fact that we limited the
output dimensions of our nonlocal boxes while measurements on
entangled states can have any number of outputs. However, we would
like to point out that the quantum universality theorem uses quantum
teleportation as its main building block, which requires
measurements with a finite set of possible outputs. We believe that
the answer might be related to the question of the difference
between entanglement measures and nonlocality measures~\cite{ms07}.
In our scenario, we do not allow the participants to use classical
communication, since it is a nonlocal resource. On the other hand,
the universality of the maximally entangled pair of qubits is
established by allowing the participants any resource that do not
increase entanglement: shared randomness, local operations
\emph{and} classical communication. If we take away that last
resource, the universality theorem of entanglement breaks down.
Therefore, nonlocality is directly used in order to generate any
possible entangled state out of a maximally entangled pair of
qubits. What does this entail precisely? We will let the reader
ponder this question.

\section*{Acknowledgements}
The authors would like to thank Hugue Blier, Gilles Brassard, Stefano Pironio, Tomer Schlank and Alain Tapp for enlightening discussions on the subject. A.A.M. is especially thankful to Gilles Brassard and the Universit\'e de Montr\'eal for the hospitality where this collaboration could be developed.  F. D. is supported by the Natural Sciences and Engineering Research Council of Canada (NSERC) through the Canada Graduate Scholarship program. N.G. and A.A.M. are supported in part by the European Commission under the Integrated Project Qubit Applications (QAP) funded by the IST directorate as Contract Number 015848 and by the Swiss NCCR \textit{Quantum Photonics}. A.H. was partially supported by an Israel Science Foundation research grant and by an Israel Ministry of Defense research grant.

\bibliographystyle{IEEEtran}
\bibliography{IEEEabrv,article}

\end{document}